\documentclass{article}
\usepackage{spconf,amsmath,graphicx}
\usepackage{tikz}
\usepackage{comment}
\usepackage{color}
\usepackage{hyperref}
\usepackage{url}
\usepackage{tabularx}
\usepackage{booktabs} 
\usepackage{caption}
\usepackage{subcaption}
\usepackage{multirow}
\usepackage[normalem]{ulem}
\usepackage{soul}

\title{Match to win: Analysing sequences lengths for efficient self-supervised learning in speech and audio}
\name{Yan Gao$^1$, Javier Fernandez-Marques$^2$, Titouan Parcollet$^3$, Pedro P. B. de Gusmao$^1$, Nicholas D. Lane$^{1,2}$}
\address{
  $^1$University of Cambridge, $^2$Samsung AI, $^3$Avignon University}

\newcommand{\oom}[1]{{\textcolor{gray}{OOM}} #1}
\copyrightnotice{978-1-6654-7189-3/22/\$31.00~\copyright2023 IEEE}

\begin{document}
%
\maketitle
\begin{abstract}
Self-supervised learning (SSL) has proven vital in speech and audio-related applications. The paradigm trains a general model on unlabeled data that can later be used to solve specific downstream tasks. This type of model is costly to train as it requires manipulating long input sequences that can only be handled by powerful centralised servers. Surprisingly, despite many attempts to increase training efficiency through model compression, the effects of truncating input sequence lengths to reduce computation have not been studied. In this paper, we provide the first empirical study of SSL pre-training for different specified sequence lengths and link this to various downstream tasks. We find that training on short sequences can dramatically reduce resource costs while retaining a satisfactory performance for all tasks. This simple one-line change would promote the migration of SSL training from data centres to user-end edge devices for more realistic and personalised applications.
\end{abstract}
\section{Introduction}
\label{sec:intro}
The success of deep learning in both speech and audio mainly relies on large-scale labelled training corpora \cite{kim2017joint,soltau2016neural,dong2018speech}. However, collecting large amounts of annotated samples is very expensive, time-consuming, and even error-prone. 
Self-supervised learning (SSL) \cite{baevski2020wav2vec,hsu2021hubert,chen2021wavlm,baevski2022data2vec,ravanelli2020multi} has emerged as a paradigm for learning robust representations from unlabelled data, which can later be used to solve specific downstream tasks such as automatic speech recognition (ASR), emotion recognition (ER), keyword spotting (KWS), and others \cite{yang2021superb}. Although SSL models achieve state-of-the-art (SOTA) performance in various applications, they are typically large in size and notoriously costly to train. For instance, a SOTA SSL model for speech and audio representation learning, such as wav2vec 2.0, requires over 7,000 GPU hours of pre-training (V100 GPUs) \cite{baevski2020wav2vec}. Such workloads force SSL models to be trained only in data centers where high-performance hardware is available and impede the use of SSL in user-end on-device personalisation based on the local data.

One of the main factors leading to large resource consumption in SSL training is the manipulation of long input sequences \cite{gao2022federated}. Current large-scale SSL models, including wav2vec 2.0 \cite{baevski2020wav2vec}, HuBERT \cite{hsu2021hubert} or WavLM \cite{chen2021wavlm} are often trained with sentences longer than 15 seconds. This dramatically increases the computing complexity at the temporal dimension and leads to long training times and higher memory costs. Several existing works \cite{chi2021audio,lee2022fithubert,lai2021parp} propose to boost training efficiency by simplifying or optimising the model architecture. However, none of them pays attention to the data-side impact of the sequence length. According to the findings in \cite{gao2022federated}, training SSL models using short sequences effectively reduces resource needs. Nevertheless, the quality of representations learned from short sequences has not yet been investigated. Additionally, it is unclear whether sequence lengths used during SSL pre-training should be directly linked to the specific type of downstream task in speech applications. For instance, KWS may require SSL pre-training using only short utterances (e.g. 1-3s), while ER or ASR tasks may need much longer sentences. Can the SSL stage be tailored to a specific subset of downstream tasks by limiting the length of the input sentences during pre-training? Moreover, would it be possible to migrate SSL model training from cloud servers to local edge devices once both compute and memory footprints have been reduced when using shorter sequences? 

In this paper, we conduct SSL pre-training on Librispeech with four handcrafted splits with specified sequence lengths ranging from 1 to 15 seconds. These length-dependent representations are then fine-tuned and evaluated on six downstream tasks over five different datasets from the SUPERB benchmark \cite{yang2021superb}. We find that the SSL representations learned from short sentence datasets (e.g. 1s) still lead to acceptable performance across all tasks. Additionally, we observe the existence of a strong link between sequence length for SSL pre-training and downstream datasets. 
These findings enable much cheaper to train SSL models that can be designed for specific downstream tasks, providing a more flexible scheme for SSL pre-training. We further investigate the concrete resource cost of SSL pre-training concerning sequence length and explore how these resource requirements (e.g., memory, compute) translate into the feasibility of training SSL models on edge devices.
We find that pre-training with short sequences would reduce memory consumption and computing complexity while obtaining a substantial acceleration when deploying models on edge devices.
Reducing resource requirements can remove the dependency on powerful hardware for SSL model training and enable previously prohibitive applications such as on-device personalised representations fine-tuning and federated self-supervised learning. Despite there being other ways to make the resource cost drop (e.g. model compression), this one-line change operation can be deployed immediately and used more often for the speech and audio community.

The main contributions of this work are: (1) We benchmark SSL pre-training with different sequence lengths on our customised datasets, and evaluate the learned representations on six downstream tasks (\S\ref{sec:setting} \& \S\ref{sec:results}). (2) We conduct a systematic analysis of resource consumption for SSL pre-training with different sentence lengths (\S\ref{sec:ssl_cost}). (3) We extend this analysis and explore the feasibility of on-device training for SSL pre-training and downstream fine-tuning (\S\ref{sec:ssl_cost} \& \S\ref{sec:ds_resource}).


\section{Background}
Self-supervised learning has proven effective in extracting high-level properties from speech and audio data, providing learned representations that achieve SOTA in various downstream tasks with minimal adaptation. Wav2vec 2.0 \cite{baevski2020wav2vec} masks the 15s length sequences of raw audio data in the latent space followed by a quantization process, where the latent representations are trained via a contrastive task. HuBERT \cite{hsu2021hubert} exploits an offline clustering phase to offer aligned target labels for a BERT-like prediction loss. 
WavLM \cite{chen2021wavlm} jointly learns masked speech prediction and denoising in pre-training.
However, all of these SOTA speech SSL models are large in size and trained on a massive number of long input sequences (more than 15s). This leads to models being intensively expensive to train and hence only being  at powerful data centres.

Only a few efforts have been yielded to boost the efficiency of speech SSL models. AUDIO ALBERT \cite{chi2021audio} proposes a lite version of the BERT-like self-supervised speech representation model, while Fithubert \cite{lee2022fithubert} obtains a thinner and deeper HuBERT model via distillation. PARP \cite{lai2021parp} conducts unstructured pruning on pre-trained wav2vec 2.0 to gain a sparsity mask followed by fine-tuning on downstream tasks while updating the mask. Many existing works have focused on increasing efficiency via model compression, however,  no work has yet considered efficiency from the input size perspective that would allow moving models from centralised servers to edge devices for better personalisation.

The recent work in \cite{gao2022federated} has taken a more aggressive step toward training SSL models in federated environments and concluded that the current experimental protocols are simply impractical for on-device training due to its massive resource consumption. However, one interesting finding from this work is that training SSL models using short sequences reduces compute costs. 

Unfortunately, no evaluation measuring the quality of the short-sentence-trained models was performed in said study. Motivated by this, we propose to benchmark SSL pre-training with different sequence lengths and evaluate them on various downstream tasks. Savings obtained by using short sentences enable people to pre-train SSL models using less powerful and less expensive hardware. Additionally, this would unleash more realistic applications that are currently constrained by efficiency issues, e.g., personalised representations fine-tuning and federated self-supervised learning.

\begin{figure}[t]
  \centering
  \includegraphics[width=0.95\linewidth]{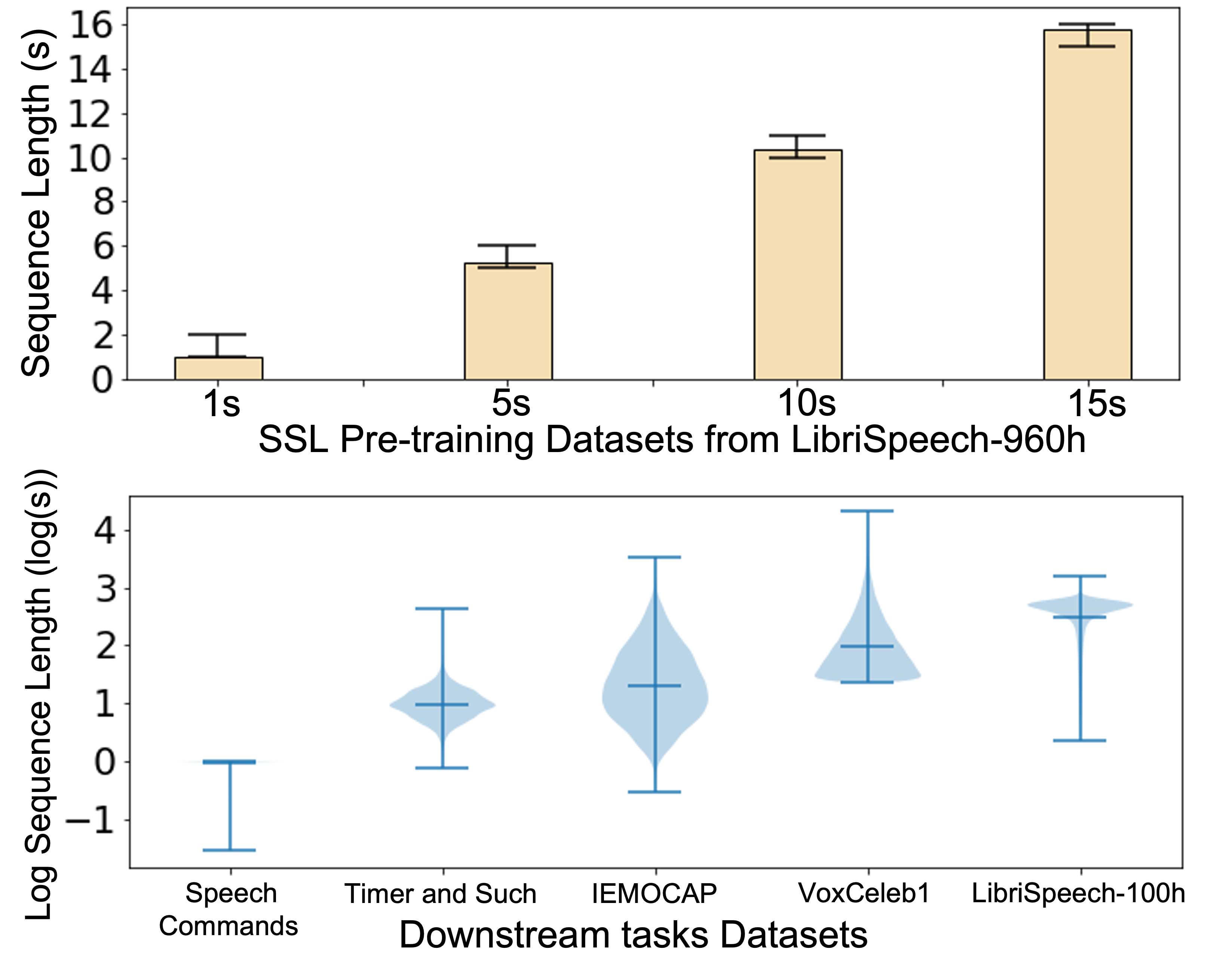}
  \caption{\small (above) The max, average and min length of four generated SSL pre-training datasets. The average lengths are 1.1s, 5.2s, 10.4s and 15.8s. (below) The distribution of log sequence length for the downstream tasks datasets. The short horizontal lines show the max, average and min log length for each dataset. The average lengths are 1.0s, 2.7s, 4.6s, 8.3s and 12.7s. \vspace{-5mm}}
  \label{fig:datasets}
\end{figure}

\section{Experimental Protocol}
\label{sec:setting}
This section describes the experimental settings for SSL pre-training on our generated datasets (\S\ref{sec:ssl_pre}), and the downstream evaluation protocols for various speech-related tasks (\S\ref{sec:downstream}).

\begin{table*}[t]
\begin{center}
    \caption{\small Evaluating wav2vec 2.0 (\textit{base}) SSL representations on various downstream tasks. The SSL models are pre-trained on the datasets with specific length of utterances ranging from 1 second to 15 seconds. KWS has two tasks with 12 or 35 commands. \textit{Froz.} represents that the SSL model is frozen at the stage of fine-tuning on downstream tasks, while \textit{Unfroz.} denotes fine-tuning the entire network. } 
    \label{tab:results}
    \scalebox{0.84}{
    \begin{tabular}{l cc|cc|cc|cc|cc|cc|cc}
    \toprule
      & \multicolumn{2}{c|}{\textbf{KWS} (12)} & \multicolumn{2}{c|}{\textbf{KWS} (35)} & \multicolumn{2}{c|}{\textbf{SF}} & \multicolumn{2}{c|}{\textbf{ER}} & \multicolumn{2}{c|}{\textbf{SID}} & \multicolumn{2}{c|}{\textbf{ASV}} & \multicolumn{2}{c}{\textbf{ASR}} \\
     \cmidrule(r){2-15}
      & \multicolumn{2}{c|}{Acc $\uparrow$} & \multicolumn{2}{c|}{Acc $\uparrow$} & \multicolumn{2}{c|}{CER $\downarrow$} & \multicolumn{2}{c|}{Acc $\uparrow$} & \multicolumn{2}{c|}{Acc $\uparrow$} & \multicolumn{2}{c|}{EER $\downarrow$} & \multicolumn{2}{c}{WER $\downarrow$} \\
      \cmidrule(r){2-15}
      \textbf{SSL models} & Froz. & Unfroz. & Froz. & Unfroz. & Froz. & Unfroz. & Froz. & Unfroz. & Froz. & Unfroz. & Froz. & Unfroz. & Froz. & Unfroz. \\
      \cmidrule(r){1-15}
      \multicolumn{1}{c}{1s} & 94.46 & 98.67 & 93.08 & 97.84 & 5.32 & 1.37 & 57.40 & 68.80 & 82.61 & 95.06 & 12.67 & 6.80 & 20.64 & 8.79 \\ 
      \multicolumn{1}{c}{5s} & 95.97 & \textbf{99.08} & \textbf{95.17} & \textbf{97.97} & \textbf{1.99} & 0.36 & \textbf{62.61} & \textbf{76.54} & 87.52 & 96.18 & 7.72 & 5.95 & 11.47 & 6.99 \\ 
      \multicolumn{1}{c}{10s} & \textbf{96.06} & 98.86 & 94.38 & 97.87 & 2.44 & 0.35 & 60.12 & 72.71 & \textbf{89.21} & \textbf{96.45} & 7.33 & 5.87 & 9.35 & 6.23 \\ 
      \multicolumn{1}{c}{15s} & 94.83 & 98.51 & 93.83 & 97.76 & 3.21 & \textbf{0.24} & 60.53 & 74.91 & 88.11 & 95.52 & \textbf{7.07} & \textbf{5.64} & \textbf{8.72} & \textbf{6.20} \\ 
     \bottomrule
     \vspace{-10mm}
    \end{tabular}
    }
\end{center}
\end{table*}

\subsection{SSL Pre-training}
\label{sec:ssl_pre}

We conduct all of the SSL models pre-training using wav2vec 2.0 \textit{base} on our generated datasets with specific utterance length from LibriSpeech \cite{7178964}. The details are as follows: 
\vspace{2mm}

\noindent \textbf{Model}. Our choice of SOTA SSL model for speech representation learning is wav2vec 2.0 \textit{base}. 
This model consists of three modules: a convolutional encoder that extracts latent acoustic features from the raw waveform; a transformer block that generates contextualised embeddings; and, a quantization module that discretizes the output of the convolutional module to a finite set of speech representations. The model is jointly trained by a contrastive loss and a diversity loss. The architecture details are the same as the HuggingFace repository \cite{wolf-etal-2020-transformers}. 
\vspace{2mm}

\noindent \textbf{Dataset}. The Librispeech corpus without transcriptions, containing 960 hours of speech from 2.5k speakers reading books, is used to generate our customized datasets for SSL model pre-training. The audio sequences are truncated or concatenated into a specified length. As a result, we produce four proposed datasets with average lengths of 1.1, 5.2, 10.4 and 15.8 seconds, while the maximum difference of the sequence lengths for each dataset is within 1 second (Fig. \ref{fig:datasets} above). The total duration of speech data (960 hours) is same for all generated datasets.  
\vspace{2mm}

\noindent \textbf{Hyper-parameters}. Following the settings in \cite{baevski2020wav2vec}, we train the SSL models with Adam \cite{kingma2014adam} on 20 V100 GPUs for 300k update steps until convergence. A warm-up learning rate scheduler is applied for the first 10\% of updates with a peak of $6 \times 10^4$, followed by a linear decay. As for the original work, the effective batch duration is set to 1.6 hours (\textit{i.e.} an optimization step is done every time we forwarded and back-propagated 1.6 hours of speech). All the training settings remain the same for the four generated datasets. 

\subsection{Downstream Evaluation}
\label{sec:downstream}
Inspired by the experimental settings in SUPERB \cite{yang2021superb}, we select six downstream tasks at various difficulty levels to evaluate the SSL models trained on the utterances with specific lengths. The learned SSL representations are either frozen or unfrozen during downstream task fine-tuning.
For a fair comparison, we adhere to hyper-parameter settings from SpeechBrain recipes \cite{ravanelli2021speechbrain}.
The details are as follows: 
\vspace{2mm}

\noindent \textbf{Keyword Spotting (KWS)} continuously listens to an audio stream to detect a preregistered set of words. The task is usually used to start an interaction with a voice assistant. Speech Commands dataset v2.0 \cite{speechcommandsv2}, consisting of 35 classes of keywords, is used for the task. We conduct KWS on these 35 classes or a subset with 12 classes including an additional class for silence and an unknown class to include the false positive. The downstream model comprises a mean-pooling followed by a linear layer with cross-entropy loss. The evaluation metric is accuracy (ACC).  
\vspace{2mm}

\noindent \textbf{Slot Filling (SF)}, as a typical task for Spoken Language Understanding (SLU), predicts a sequence of semantic slot-types from an utterance. The slot-type labels are represented as special tokens to wrap the slot-values in transcriptions. We choose \emph{Timers and Such} dataset \cite{lugosch2021timers} for this task. The dataset consists of spoken English commands for common voice control use cases involving numbers. We use train-real (1,640 audios) and train-synth (192,000 audios) sets for fine-tuning while test-real set (240 audios) for evaluation. As for the downstream model, a 2-layer 1024-unit Bidirectional LSTM is adopted and optimized by CTC loss. The evaluation metric is the character error rate (CER).   
\vspace{2mm}

\noindent \textbf{Emotion Recognition (ER)} predicts an emotion class for the input utterances. We select the widely used ER dataset IEMOCAP \cite{busso2008iemocap} for the task. Following the conventional evaluation protocol, we only use the four main classes (anger, happiness, sadness, and neutral) by discarding the unbalance emotion classes and cross-validate on five folds of the leave-two-speaker-out splits. Similar to KWS, a mean-pooling followed by a linear transformation is adopted for the downstream model, optimised with cross-entropy loss. The evaluation metric is accuracy (ACC).   
\vspace{2mm}

\noindent \textbf{Speaker Identification (SID)} predicts the speaker identity for a given utterance, and the same speakers appear in both training and test sets. VoxCeleb1 dataset \cite{nagrani2020voxceleb}, containing 350 hours of audio from 1211 speakers, is used for this task. For the downstream model, we adopt x-vector \cite{snyder2018x} trained with cross-entropy loss. The accuracy (ACC) is reported on the validation set.  
\vspace{2mm}

\noindent \textbf{Automatic Speaker Verification (ASV)} is used to determine whether the speakers match a pair of utterances as a binary classification. Speakers in the test set do not appear in the training set, leading to a more challenging task than SID. We also choose VoxCeleb1 \cite{nagrani2020voxceleb} for the task and adopt x-vector for speaker embedding training. Then, the cosine-similarity is used to produce pairwise matching scores. The evaluation metric is equal error rate (EER).  
\vspace{2mm}

\noindent \textbf{Automatic Speech Recognition (ASR)} predicts the transcription of the input utterances by mapping acoustic data into words, which serves as the most challenging downstream task. LibriSpeech \cite{7178964} train-clean-100/dev-clean/test-clean subsets are used for training/validation/test. We use a 2-layer 1024-unit Bidirectional LSTM as a downstream model trained with CTC loss at character level.
The evaluation metric is word error rate (WER).  
\vspace{2mm}

Regarding the sequence length for the downstream task datasets (Fig. \ref{fig:datasets} below), Speech Commands and Timer-and-such only contain short utterances with average length 0.98 and 2.73 seconds. IEMOCAP as an ER dataset has longer sentences (average 4.55 seconds). VoxCeleb1 shows a high variance of sequence length, but most of the utterances are below 10 seconds. LibriSpeech-100h has the largest average sequence length (12.69 seconds) corresponding to the most challenging downstream task ASR.

\begin{figure}[t]
  \centering
  \includegraphics[width=\linewidth]{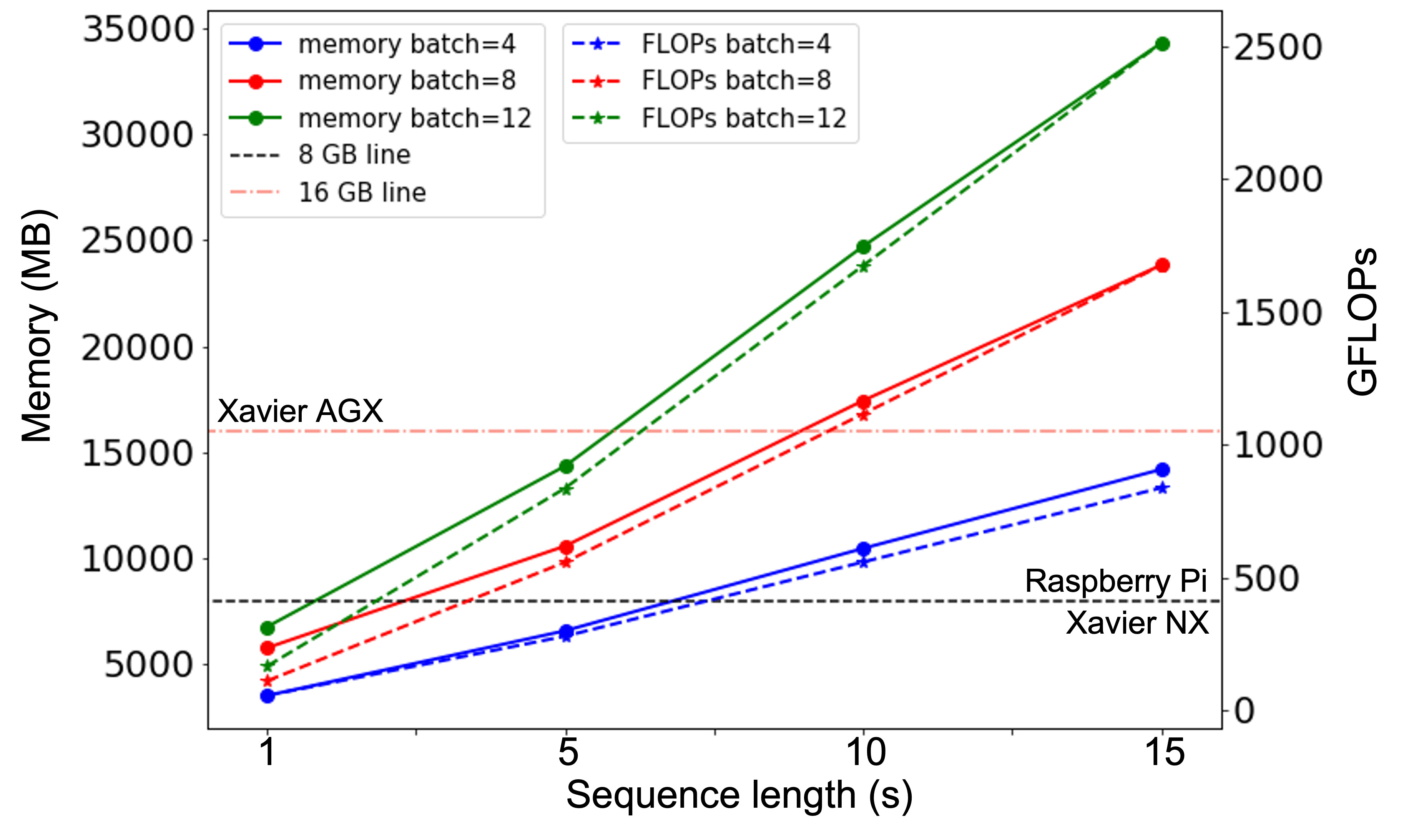}
  \caption{\small Memory utilization and FLOPs of wav2vec 2.0 \textit{base} model with different sequence length and batch size. The memory and FLOPs grow linearly with the length of the input sequence. \vspace{-5mm}}
  \label{fig:mem}
\end{figure}

\section{Experimental Results}
\label{sec:results}

This section shows the downstream performance for pre-trained SSL models and analyses the correlation between sequence lengths of SSL pre-training and downstream datasets.  
\vspace{2mm}

\noindent \textbf{Downstream Performance}. Table \ref{tab:results} shows the performance of wav2vec 2.0 (\textit{base}) SSL representations on various downstream tasks. First, all of the SSL models trained on different sequence lengths successfully learn meaningful speech representations, albeit they exhibit some performance differences on the downstream tasks. 
Second, fine-tuning the entire network obtains higher performance than freezing the SSL model at the phase of downstream training. For instance, the best unfrozen models exceed the frozen ones by 3.02\%, 2.80\%, 1.75\%, 13.93\%, 7.24\%, 1.43\%, 2.52\% for the tasks of KWS (12), KWS (35), SR, ER, SID, ASV, and ASR, respectively.
This is expected as more trainable parameters can participate in the fine-tuning. However, this leads to higher resource consumption (see \S\ref{sec:ds_resource}). 
Third, the best results of frozen and unfrozen settings are from different SSL models for the task of KWS (12) and SF, while other tasks are consistent with respect to the frozen and unfrozen cases. We argue that the freezing setup serves as the direct indication of representations’ quality as the learned SSL model remains static during the fine-tuning on the downstream tasks.

When comparing the frozen SSL models pre-trained on different lengths of utterances, the ``1s" model has the worst performance on all tasks, but it still achieves 93.1\%, 94.5\% accuracy on KWS (12) and KWS (35) compared to 95.2\%, 96.0\% for the best models. 
The performance gap between the ``1s" model and the others trained on longer sequences is larger on other tasks, especially for ASR which obtains 11.92\% higher WER than the best model trained with 15s sequences.
This indicates that very short utterances are insufficient for learning robust speech representations since the short input sequence would not be able to provide sufficient spatiotemporal information for representation learning (critical for certain downstream tasks).
Interestingly, the longest sentences (i.e. 15s) do not perform best on all downstream tasks. For instance, the frozen ``5s" SSL model gains the best performance on KWS (35), SF and ER.
Additionally, the ``10s" model performs best on SID, while the ``15s" model obtains the lowest error rate on ASV and ASR tasks.  
\vspace{2mm}

\begin{table*}[t]
\begin{center}
    \small
    \caption{\small Training time (average seconds per batch) of wav2vec 2.0 \textit{base} model with different sequence lengths and batch size. OOM stands for out of memory. The medium-tier GPU (RTX 2080Ti) with 11GB of memory can train with the sequences of at most 10s length. Edge devices with 8GB of memory are significantly constrained in terms of sequence length and batch size. In practical terms, these devices can only train with shorter 5s-long sequences. With the AGX, which doubles memory, one could either increase the batch size (from 4 to 8) or sequence length (from 5s to 10s). Overall, the main limitation is memory since the GPU-enabled devices have sufficient compute to train this medium-sized SSL model.}
    \label{tab:ssl_time}
    \scalebox{1.0}{
    \begin{tabular}{lrcccccccccc}
    \toprule
     & & \multicolumn{2}{c}{Batch=1} & \multicolumn{2}{c}{Batch=2} & \multicolumn{2}{c}{Batch=4} & \multicolumn{2}{c}{Batch=8} & \multicolumn{2}{c}{Batch=12} \\
   \cmidrule(r){3-4} \cmidrule(r){5-6} \cmidrule(r){7-8} \cmidrule(r){9-10} \cmidrule(r){11-12}
    \textbf{Device} & \textbf{Length} &  FP32 & Mixed & FP32 & Mixed & FP32 & Mixed & FP32 & Mixed & FP32 & Mixed \\
     \cmidrule(r){1-12}
      \multirow{4}{*}{\parbox{2.87cm}{\textbf{NVIDIA RTX 2080Ti} (11GB - 6CPU+GPU)}} & 1s & 0.10 & 0.09 & 0.10 & 0.09 & 0.10 & 0.09 & 0.12  & 0.10 & 0.12 & 0.11  \\
      & 5s & 0.11 & 0.09 & 0.12 & 0.11 & 0.17 & 0.14 & 0.28  & 0.23 & \oom & 0.26   \\
      & 10s & 0.12 & 0.11 & 0.17 & 0.16 & 0.30 & 0.24 & \oom  & \oom  & \oom & \oom  \\
      & 15s & 0.16 & 0.14 & 0.26 & 0.23 & \oom & \oom & \oom  & \oom & \oom & \oom    \\
      \cmidrule(r){1-12}
      \multirow{4}{*}{\parbox{2.1cm}{\textbf{Raspberry Pi 4} (8GB - CPU)}} & 1s & 8.33 & — & 10.33 & — & 14.80 & — & 24.59  & — & 31.65 & —  \\
       & 5s & 18.81 & — & 32.16 & — & 61.59 & — & \oom  & — & \oom & —   \\
       & 10s & 35.00 & — & 65.26 & — & \oom & — & \oom  & —  & \oom & —  \\
       & 15s & 45.56 & — & \oom & — & \oom & — & \oom  & — & \oom & —    \\
      \cmidrule(r){1-12}
       \multirow{4}{*}{\parbox{2.55cm}{\textbf{Jetson Xavier NX} (8GB - CPU+GPU)}} & 1s & 0.40 & 0.35 & 0.38 & 0.36 & 0.52 & 0.46 & 0.88 & 0.82 & 1.24 & 0.82  \\
      & 5s & 0.70 & 0.59 & 0.99 & 0.92 & 1.90 & 1.83 & \oom  & \oom & \oom & \oom \\
      & 10s & 1.19 & 1.05 & \oom\textsuperscript{\hspace{-1mm}$\dagger$} & 1.80 & \oom & \oom & \oom  & \oom & \oom &  \oom  \\
      & 15s & 1.59 & 1.58 & \oom & \oom & \oom & \oom & \oom  & \oom & \oom & \oom   \\
      \cmidrule(r){1-12}
      \multirow{4}{*}{\parbox{2.75cm}{\textbf{Jetson Xavier AGX} (16GB -  CPU+GPU)}} & 1s & 0.24 & 0.19 & 0.23 & 0.23 & 0.32 & 0.29 & 0.49 & 0.35 & 0.60 & 0.43  \\
      & 5s & 0.38 & 0.36 & 0.59 & 0.55 & 1.05 & 0.89 & 1.84  & 1.26 & \oom & \oom \\
      & 10s & 0.62 & 0.61 & 1.09 & 0.97 & 2.07 & 1.64 & \oom  & \oom & \oom & \oom   \\
      & 15s & 0.94 & 0.88 & 1.69 & 1.44 & \oom & \oom & \oom  & \oom & \oom & \oom   \\
     \bottomrule
     \addlinespace[2pt]
    \multicolumn{12}{l}{\scriptsize $\dagger$ Despite Jetson NX and RPi4 having the same memory (8GB), slighlty less RAM is available for applications in the former.}\\
    \vspace{-10mm}
    \end{tabular}
    }
\end{center}
\end{table*}

\noindent \textbf{Lengths Matching}. From Table \ref{tab:results}, we find that the sequence length on SSL pre-training correlates with the length of the best performing downstream model for each dataset. For instance, the frozen SSL model trained with 5s sequences performs best on SF and ER, where the corresponding downstream datasets have an average sentence length of 2.7s and 4.6s. Once the downstream average sequence length increases to 8.3s and 12.7s for SID/ASV and ASR, the best SSL models are from 10s and 15s pre-training datasets. 
The KWS tasks are an exception as the best-performed models are from 5s or 10s pre-training datasets. This can be explained by: (1) sequences with 1s length are too short to extract sufficient information during SSL training; (2) KWS is a simple multi-class classification task, and as a result, it is hard to see much difference between different pre-training settings.
As a result, we argue there exists a strong correlation between sequence lengths in SSL pre-training and downstream datasets, i.e., matching sequence lengths for SSL pre-training to those needed for the downstream task would yield the best performing models. 

However, the existing work \cite{baevski2020wav2vec,hsu2021hubert,chen2021wavlm} only uses 15s sequence lengths for SSL pre-training, which would not perform best for all downstream tasks. Additionally, training with longer sequences will cause dramatically higher resource consumption, mainly due to the contextual information captured by the transformer module in wav2vec 2.0 model. This aspect is analysed in detail in the following section.

\section{System Resource Analysis}
We present measurements of resource consumption for SSL pre-training with different lengths of sequences (\S\ref{sec:ssl_cost}) and downstream fine-tuning with various tasks (\S\ref{sec:ds_resource}), while promoting both stages into on-device training.

\subsection{Resource Cost for SSL Pre-training}
\label{sec:ssl_cost}
Training SSL model with shorter sequences would reduce the resource cost and hence enable the possibility of on-device training, allowing the devices to fine-tune personalised representations based on the local data. 
This section provides a quantitative analysis in terms of the memory consumption and computational complexity of wav2vec 2.0 \textit{base} model with respect to sequence length, followed by an exploration of training such model on various edge devices. 
\vspace{2mm}

\noindent \textbf{Memory cost and FLOPs}. Figure \ref{fig:mem} shows the memory utilization and floating point operations (FLOPs) with different input sequence lengths and batch sizes. An immediate observation is that memory and FLOPs grow linearly with the length of the input sequence. For instance, the memory cost with 15 seconds sentences is around 4$\times$, 4$\times$, 5$\times$ higher than the one with one-second utterances when batch size equals to 4, 8, 12, respectively. The gap in FLOPs is more significant with approximately 15$\times$ higher computing complexity when comparing 15 and 1 seconds length sequences.

As the discussion in \S\ref{sec:results}, there exists a strong correlation of sequence lengths between SSL pre-training and downstream datasets. However, currently released large-scale SSL models including wav2vec 2.0, HuBERT \cite{hsu2021hubert} or WavLM \cite{chen2021wavlm} are often trained with sentences longer than 15s. Here, the SSL stage could be tailored to a specific subset of downstream tasks and limit the length of the input sentences for pre-training, reducing the compute and memory footprints of such process. This enables SSL models to able to train on medium-tier GPU (Table \ref{tab:ssl_time}), and even on the edge devices.
\vspace{2mm}

\noindent \textbf{On-device Training Feasibility}. Adjusting sequence length and batch size could enable SSL pre-training to migrate from a centralised server to local devices, hence leading to federated representations learning based on the users' data. In practice, memory is a scarce resource in edge devices and limits the feasibility of on-device training. For instance, some popular mobile processors and edge devices including the Raspberry Pi 4 (RPi) or the Jetson Xavier NX are only accompanied with 8GB of memory, while the higher-end Jetson Xavier AGX comes with 16GB. According to Fig. \ref{fig:mem}, only the latter could train the SSL model with sequences of 10 or 15 seconds on batch size 4. However, once increasing batch size to 8 or 12, a typical case with current SSL models, none of the devices may be able to train the model. Nevertheless, with sequences length down to 5 seconds, even RPi and Xavier NX could afford the memory cost on 4 batch size. While training with 1s sequences, the edge devices can allow a larger batch size of up to 12. However, we should note that these devices would likely be running other applications and processes in the background. As a result, the memory ceiling available for model training workloads is lower than the devices' total memory. Therefore, it is necessary to show the real utilization by deploying the models on these devices.

\begin{table}[t]
\begin{center}
    \caption{\small Training time (average seconds per batch) of different downstream tasks with frozen or unfrozen wav2vec 2.0 \textit{base} model. The sequence length for each task is the average of its corresponding dataset (Fig. \ref{fig:datasets}). B represents the batch size, while OOM stands for out-of-memory. Hyper-parameters remain default following SpeechBrain \cite{ravanelli2021speechbrain} recipes.} 
    \label{tab:ds_time}
    \scalebox{0.8}{
    \begin{tabular}{lrrr|rr|rr}
    \toprule
      & & \multicolumn{2}{c|}{\textbf{Raspberry Pi}} & \multicolumn{2}{c|}{\textbf{Jetson NX}} & \multicolumn{2}{c}{\textbf{Jetson AGX}} \\
     \cmidrule(r){3-8}
      \textbf{Tasks} & \textbf{B} & Froz. & Unfroz. & Froz. & Unfroz. & Froz. & Unfroz.   \\
      \cmidrule(r){1-8}
      KWS & 32 & 47.16 & 167.49 & 0.70 & 2.89 & 0.42 & 1.70  \\ 
      SF & 8 & 213.98 & 239.84 & 3.67 & 5.40 & 2.28 & 2.93  \\ 
      ER & 4 & 28.98 & 125.23 & 0.51 & 2.08 & 0.30 & 1.23  \\ 
      SID/ASV & 8 & 18.85 & 92.89 & 0.65 & 2.39 & 0.38 & 1.44  \\ 
      ASR & 6 & \oom & \oom & \oom & \oom & 2.98 & 6.57  \\ 
     \bottomrule
     \vspace{-12mm}
    \end{tabular}
    }
\end{center}
\end{table}

In Table \ref{tab:ssl_time}, we benchmark the observed training time (i.e. seconds per batch) of wav2vec 2.0 \textit{base} model with different sequence lengths and batch sizes on the aforementioned devices. First, and following the observation in Fig. \ref{fig:mem}, all devices run into OOM errors when processing sequences longer than 10s and with batch size larger than 8 sequences. In other words, it is intractable for on-device SSL pre-training to follow the standard experimental protocol (i.e.15s). The higher-end Xavier AGX is able to train the model with 10s sequences once the batch size decreasing to 4, and enables handling even 15s if using batch size 2. 
Parameter precision is another element that affects the memory and compute footprint of an architecture. Therefore, we compare the impact of full-precision {\tt FP32} with mixed precision ({\tt FP32}/{\tt FP16}) \cite{micikevicius2018mixed}. Mixed precision provides a substantial boost in training speed for devices supporting it: at most 1.51$\times$ and 1.46$\times$ faster than full precision for the Xavier NX and AGX.

From the previous analysis, it would be practical to conduct SSL pre-training locally on edge devices if acceptable downstream tasks performance can be assured. Training with extremely short sequences (e.g. 1s) would largely preserve computing and memory resources, however the downstream performance would likely drop if it is designed to operate with longer sequences. This provides a trade-off between the length selection and final evaluation performance. Additionally, although these devices typically have lower computing capability than server-grade GPUs, training could be parallelized on hundreds of devices following a federated learning framework design. This would enable not only performing training on potentially much larger amounts of data but also would ensure that data remains private, i.e., without leaving the user's device.

\begin{figure}[t]
  \centering
  \includegraphics[width=1.0\linewidth]{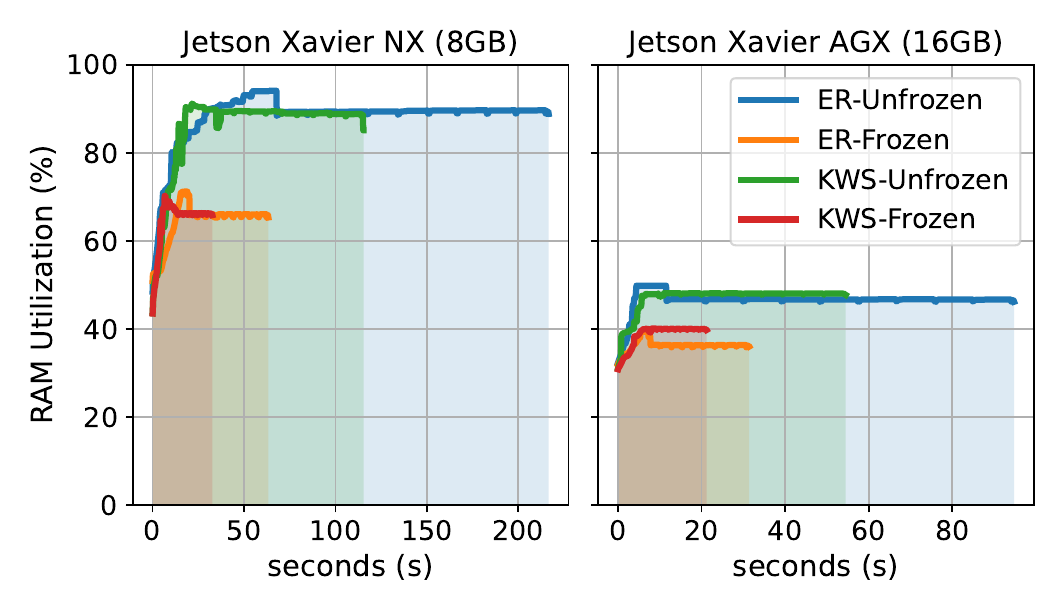}
  \caption{\small Memory footprint when performing ten epochs of on-device training for KWS and ER downstream tasks on Jetson NX and AGX. The memory footprint as well as wall-clock is much lower in settings with a frozen wav2vec2.0. If memory is limited, utilizing a frozen SSL backbone would be the only way to ensure other applications can run concurrently during on-device downstream task learning. \vspace{-3mm}}
  \label{fig:pl}
\end{figure}

\subsection{Resource Cost for Downstream Tasks}
\label{sec:ds_resource}
 A natural next step is to explore the feasibility of on-device training for downstream models on the devices previously considered for SLL pre-training. This would facilitate the personalised fine-tuning based on users' data for different downstream applications.

Table \ref{tab:ds_time} shows the per-batch training time with the sequences of average length for each downstream dataset. One can see that all downstream tasks except ASR could be trained on all three devices. The computational complexity for different tasks is not only from the input sequence length but also a function of batch size and downstream model architectures. This leads to the variation of training time on the devices. Particularly, training with a frozen SSL model is faster than fine-tuning the entire network. For instance, it provides 4.0$\times$, 1.3$\times$, 4.1$\times$, 3.8$\times$ and 2.2$\times$ acceleration on Xavier AGX for the task of KWS, SF, ER, SID/ASV and ASR, respectively. Along the same lines, Figure~\ref{fig:pl} shows that the memory footprint of frozen SSL backbones is much lower.


\section{Conclusion}
In this paper, we provided the first empirical study of SSL pre-training using application-dependent sequence lengths and linked it to various downstream tasks. Based on our systematic analysis of resource costs, our results indicate that the overhead of SSL pre-training can be significantly reduced if sequence length is tailored to specific types of downstream tasks and, in particular, when these rely on short sequence lengths. However, for being able to apply the SSL model to any downstream task, training on longer sequences remains the best setup.
This sets the foundations for migrating SSL training to local edge devices in more realistic speech and audio-related applications.

\section{Acknowledgment}
This work was performed using HPC/AI resources from GENCI-IDRIS (Grant 2021-A0111012991).



\bibliographystyle{IEEEbib}
\bibliography{strings,refs}

\end{document}